\documentclass[12pt]{iopart}

\usepackage{setstack}
\usepackage{iopams}

\usepackage{graphicx}
\usepackage[colorlinks]{hyperref}
\usepackage{siunitx}
\usepackage{multirow}
\usepackage[Export]{adjustbox}
\adjustboxset{max width=\linewidth}

\begin{document}

\newcommand*{\Ca}{\text{$^{40}$Ca$^+$}}
\newcommand*{\Sonehalf}{\text{4S$_{1/2}$}}
\newcommand*{\Ponehalf}{\text{4P$_{1/2}$}}
\newcommand*{\Pthreehalf}{\text{4P$_{3/2}$}}
\newcommand*{\Dthreehalf}{\text{3D$_{3/2}$}}
\newcommand*{\Dfivehalf}{\text{3D$_{5/2}$}}
\newcommand*{\trans}[2]{\text{#1~$\leftrightarrow$~#2}}

%

\providecommand*{\ket}[1]{\left|#1\right\rangle}
\providecommand*{\bra}[1]{\left\langle#1\right|}
\providecommand*{\order}[1]{\mathcal{O}(#1)}
\providecommand*{\Tr}{\text{Tr}}


\title{Ultrafast coherent excitation of a \Ca{} ion}



\author{D Heinrich, M Guggemos, M Guevara-Bertsch, M I Hussain, C Roos and R Blatt}
\address{Institut f\"ur Quantenoptik und Quanteninformation,
\"Osterreichische Akademie der Wissenschaften, Technikerstr. 21a,
6020 Innsbruck, Austria}
\address{Institut f\"ur Experimentalphysik, Universit\"at Innsbruck,
Technikerstr. 25, 6020 Innsbruck, Austria}

\ead{daniel.heinrich@uibk.ac.at}


\date{\today}

\begin{abstract}

Trapped ions are a well-studied and promising system for the realization
of a scalable quantum computer.
Faster quantum gates would greatly improve the applicability of such a system
and allow for greater flexibility in the number of calculation steps.
In this paper we present a pulsed laser system, delivering picosecond pulses at a
repetition rate of \SI{5}{\GHz} and resonant to the
\trans{\Sonehalf}{\Pthreehalf} transition in \Ca{} for coherent population
transfer to implement fast phase gate operations.
The optical pulse train is derived from a mode-locked, stabilized optical
frequency comb and inherits its frequency stability.
Using a single trapped ion, we implement three different techniques for measuring
the ion-laser coupling strength and characterizing the pulse train emitted by the
laser,
and show how all requirements can be met for an implementation
of a fast phase gate operation.
\end{abstract}




\maketitle

\section{Introduction}
  \label{sec:intro}

Trapped ions are a promising system for the implementation of
a scalable quantum computer
\cite{Cirac1995,Kielpinski2002,Wineland2003,Garcia-Ripoll2005a,Haffner2008}.
Two-qubit entangling gate operations
have been demonstrated \cite{Schmidt-Kaler2003,Leibfried2003,Ballance2016,Gaebler2016}
and combined with single-qubit gates to build an elementary quantum processor 
\cite{Schmidt-Kaler2003a,Schindler2013,Debnath2016}.
The entangling gate operations in these experiments rely on
spectroscopically-resolved motional sidebands of the ion crystals,
a requirement that limits the duration of a gate operation to more
than the period of motion of the ions in the trap
(typically a few $\si{\micro\s}$ or more).
Overcoming this limitation would advance the development of a
scalable quantum computer as it
would allow one to increase the number of gate operations
(computational steps) that can be completed
within the coherence time of the ion-qubits.

Two-qubit entangling gate operations in less than one trap period
have been proposed by Garc\'{i}a-Ripoll, Zoller and Cirac in 2003 
\cite{Garcia-Ripoll2003} using counter-propagating laser pulses.
Several groups are working on its realization
\cite{Mizrahi2014,Hussain2016}
but so far only single-qubit gate operations
\cite{Madsen2006,Campbell2010} and single-ion
spin-motion entanglement \cite{Mizrahi2013} have been reported
on time scales shorter than the ion oscillation period.
Recently, creation of two-qubit entanglement by a train of ultrafast laser
pulses within a few microseconds has been demonstrated in the ground-states
of a pair of Yb$^+$ ions \cite{Wong-Campos2017}.

Our goal, beyond the scope of this work, is to implement an ultrafast
two-qubit phase gate operation \cite{Garcia-Ripoll2003,Taylor2017}
using resonant, counter-propagating laser pulses
and to complete it in less than one trap period.
The scheme uses pairs of $\pi$-pulses
for applying ion-state dependent momentum kicks to a two-ion crystal.

\begin{figure}
\centering
 \includegraphics[width=0.8\textwidth]{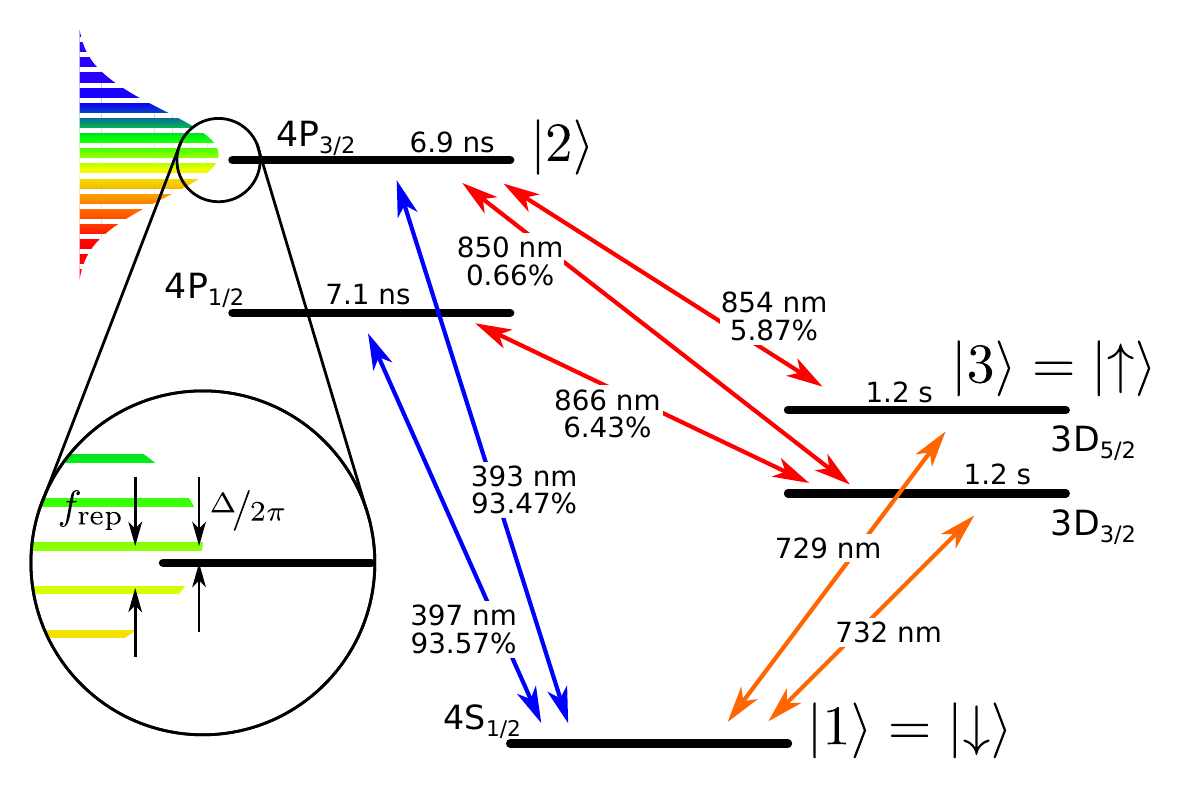}
 \caption{
 Energy level scheme of \Ca showing the levels and transitions
 relevant to the experiment.
 Possible transitions are shown with their
 wavelengths and branching ratios \cite{Gerritsma2008,Ramm2013},
 excited electronic states with their lifetimes \cite{Jin1993,Barton2000,Kreuter2005}.
 The bell curve-shaped bars in the upper-left corner are a representation of
 the pulsed laser's spectral modes.
 }
 \label{fig:levelscheme}
 \label{fig:simplelevelscheme}
\end{figure}

Figure \ref{fig:simplelevelscheme} shows a level scheme of \Ca{}
with the levels relevant to the gate operation.
We encode each qubit in two Zeeman substates of the \Sonehalf{}
($= \ket{\downarrow}$) and \Dfivehalf{} ($= \ket{\uparrow}$)
states of a \Ca{} ion \cite{Schmidt-Kaler2003a}.
The laser pulses resonantly excite the \trans{\Sonehalf}{\Pthreehalf}
transition \cite{Bentley2015}.
Depending on the qubits' state, each ion will either
not interact with the pulses at all (qubit state $\ket{\uparrow}$)
or absorb one photon from the first pulse of each pair and emit a photon
into the second pulse (qubit state $\ket{\downarrow}$),
gaining a momentum of $2 \hbar \overrightarrow{k}$ in the direction of the first pulse,
where $\overrightarrow{k}$ is the photons' wave vector.
By subjecting the ion crystal to a sequence of momentum kicks and times of
free evolution of the crystal in the trap potential,
the ions are forced to follow different, state-dependent
trajectories through phase-space.
The area enclosed by these trajectories corresponds to the phase the state
acquires during the pulse sequence \cite{Leibfried2003}.
When the relative phase between the state-pairs $\ket{00}$, $\ket{11}$ and
$\ket{01}$, $\ket{10}$ is $\pi/2$ and
the pulse sequence returns both the center-of-mass and the breathing
mode of motion \cite{James1998} of the ion crystal to the initial state,
the operation will be the desired geometric phase gate \cite{Garcia-Ripoll2003}.
Both conditions can be met by carefully choosing the duration of the times
of free evolution in the pulse sequence.
In order to complete such a phase gate within one trap period,
the rate $f_\pi$ at which pairs of $\pi$-pulses
are applied to the ions must be much larger than the trap frequency $\nu$
\cite{Bentley2015}.
In general, the larger $f_\pi$, the faster the gate operation can be completed.

In order for the laser system to be suitable to implement such a fast gate operation,
it should satisfy four requirements:
(1) The system needs to provide pulsed laser light with
a repetition rate much larger than the trap frequency of \SI{\sim 1}{\MHz},
in order to provide us with fine-grained control over the timing of pulse sequences.
(2) The pulse length $\delta t$ has to be much shorter than the \Pthreehalf{} state's
lifetime of \SI{6.9}{\ns} \cite{Jin1993}
to avoid or at least minimize spontaneous decay.
(3) The center frequency has to be resonant with the \trans{\Sonehalf}{\Pthreehalf}
transition at \SI{393.366}{\nm} and
(4) the laser needs to have an intensity such that for every laser pulse
$\Omega \, \delta t = \pi$, where $\Omega$ is the corresponding Rabi frequency.
Alternatively to (3), non-resonant pulses can be used to apply state-dependent
momentum kicks \cite{Campbell2010}.
In order to realize this alternative, we want to be able to tune the laser in between the two fine
structure components \trans{\Sonehalf}{\Ponehalf} and
\trans{\Sonehalf}{\Pthreehalf} such that Stark shifts cancel,
while limiting overlap of the laser's optical spectrum with either of the
two transition frequencies.

In this paper we describe and characterize a laser system which was designed
to generate the light pulses for ultrafast quantum gate operations.
Our laser system satisfies requirements (1) to (3), and
we will show below, that the intensity of the laser is
sufficiently high for generating pulses that flip the state of the ion
with \SI{96.4+-1.9}{\%} probability (requirement 4).
In order to do that, we present and characterize three methods to extract information
on the Rabi frequency
and compare them in terms of applicability and prerequisites.
The methods allow us to measure the rotation angle per pulse $\theta = \Omega \, \delta t$ and
are more accurate than simply deducing the Rabi frequency from a
measurement of the laser's optical intensity at the point of the ion \cite{Farrell1995,James1998}.

The paper is structured as follows:
Section \ref{sec:lasersystem} describes our laser setup in detail.
In section \ref{sec:theory} we introduce two methods that we developed to gain
information on $\theta$ by carrying out measurements on a trapped \Ca{} ion,
and we compare them to measuring $\theta$ by
injecting single pulses between the two Ramsey zones of a Ramsey experiment
\cite{Madsen2006}.

\section{Laser System}
  \label{sec:lasersystem}

\begin{figure}
\centering
 \includegraphics{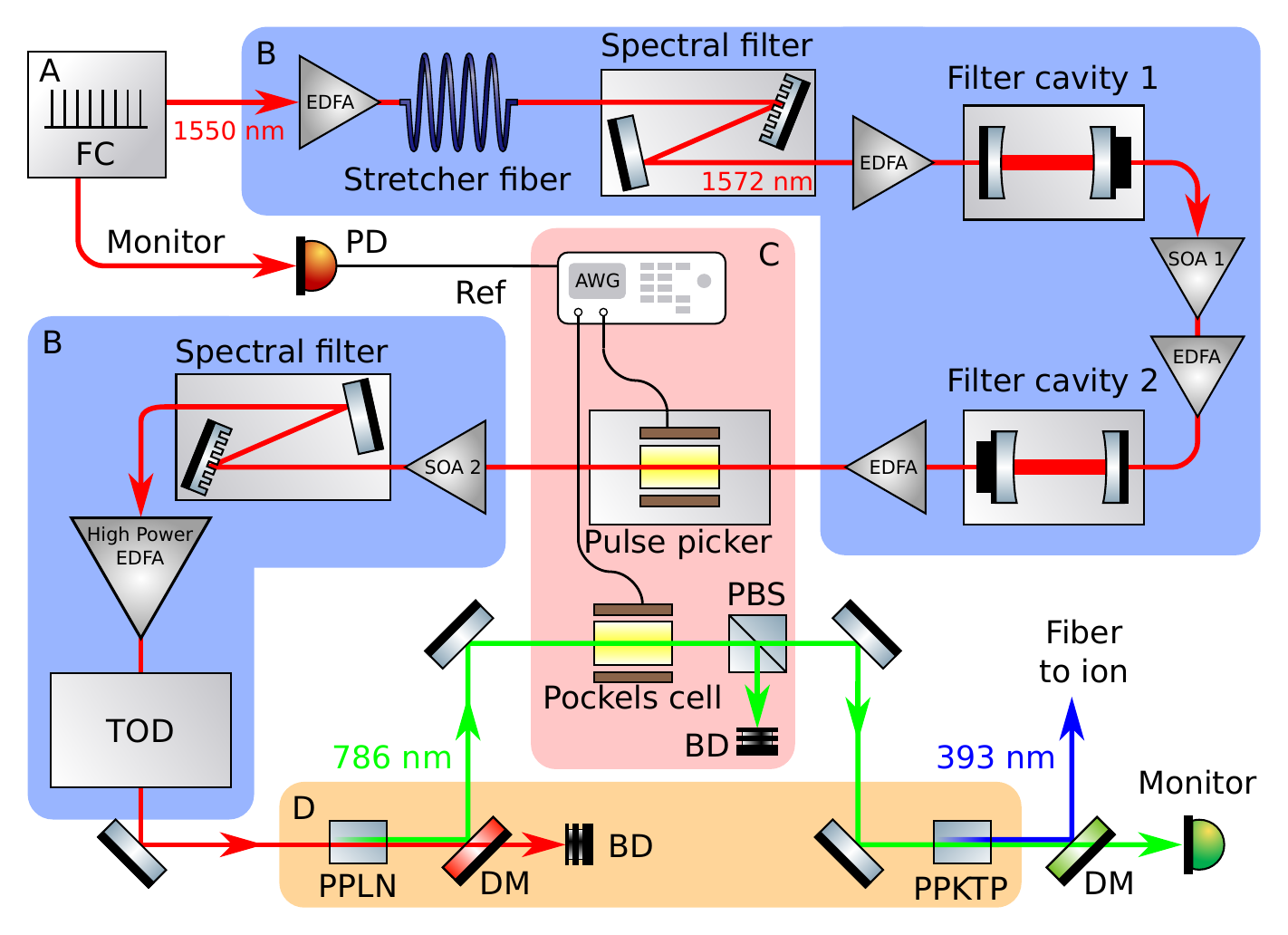}
 \caption{
  Schematic setup of our laser system.
  Panels marked A, B, C, D are discussed in subsections \ref{sec:seed},
  \ref{sec:manipulation}, \ref{sec:picking} and \ref{sec:shg}, respectively.
  AWG: arbitrary waveform generator,
  BD: Beam dump,
  DM: dichroic mirror,
  EDFA: erbium-doped fiber amplifier,
  FC: optical frequency comb,
  PBS: polarizing beam splitter,
  PD: photo detector,
  PPLN: periodically poled lithium niobate,
  PPKTP: periodically poled potassium titanyl phosphate,
  SOA: Semiconductor optical amplifier,
  TOD: third order dispersion compressor.
 }
 \label{fig:setup}
\end{figure}

Picosecond laser systems with repetition rates on the order of \si{\GHz}
and a center wavelength of \SI{393.366}{\nm} can be constructed by
frequency-quadrupling the light generated by commercial lasers 
operating in the L-band of optical fiber communication 
(\SIrange{1565}{1625}{\nm}).
An overview of the optical setup is provided in figure \ref{fig:setup}.
The system is seeded by an optical frequency comb, which is described in
more detail in subsection \ref{sec:seed}.
Filter cavities serve to multiply the laser's repetition rate.
Subsequently, the desired laser wavelength range is selected by spectral
filters and the laser output is amplified (subsection \ref{sec:manipulation}).
Fast and slow pulse picking elements enable the selection of arbitrary pulse
sequences (subsection \ref{sec:picking}).
Finally, the laser frequency is quadrupled by two single-pass second-harmonic
generation stages (subsection \ref{sec:shg}).
Measurements of the basic pulse characteristics are presented in subsection
\ref{sec:pulseswitching}.

\subsection{Seed laser}
  \label{sec:seed}

The seed laser is a fiber-based optical frequency comb
(Menlo Systems FC1500-250-WG) with a repetition rate of \SI{250}{\MHz}.
The mode locked laser creates short pulses of \SI{70}{\fs} pulse width with a
center wavelength of \SI{1550}{\nm} and a spectral bandwidth of \SI{50}{\nm}.
We lock both the carrier envelope offset and the
repetition rate to a frequency reference provided by a GPS-disciplined
oven-controlled crystal oscillator (Menlo Systems GPS 6-12) with a
fractional frequency instability of \num{\sim 2e-13}.

\subsection{Pulse manipulation}
  \label{sec:manipulation}

The pulses produced by the frequency comb need to be amplified and
frequency-upconverted in order to create $\pi$-pulses on the ions.
Furthermore, the spectral bandwidth should be limited to about \SI{1}{\THz}
($\overset{\scriptscriptstyle\wedge}{=}\SI{0.5}{\nm}$ at \SI{393}{\nm})
to avoid residual off-resonant excitation
of the \trans{\Sonehalf}{\Ponehalf} transition, and
the center wavelength should be resonant to the \trans{\Sonehalf}{\Pthreehalf}
transition.
In order to have a higher resolution of the pulse timings,
we additionally multiply the repetition rate by a factor of 20.

Erbium doped fiber (pre-)amplifiers (EDFAs) and semiconductor optical amplifiers
(SOAs) are used to compensate insertion loss at various
stages in the set-up (panels B in figure \ref{fig:setup}).
After the first preamplifier, the pulse train travels through a stretcher
fiber which adds dispersion and stretches the pulses from \SI{70}{\fs} to
\SI{\sim 50}{\ps} for chirped pulse amplification \cite{Strickland1985}.
The pulse train then travels through a spectral filter which selects the new
center wavelength of $4 \cdot \SI{393}{\nm} = \SI{1572}{\nm}$ and reduces the
spectral bandwidth to \SI{8}{\nm}.
Two filter cavities with a free spectral range of \SI{5}{\GHz} each then
increase the repetition rate from \SI{250}{\MHz} to \SI{5}{\GHz} by
transmitting only the light's spectral modes that are \SI{5}{\GHz} apart and
suppressing all others.
The second cavity's purpose is to increase the extinction ratio and therefore
equalize optical intensity of the output pulses.
The subsequent fast pulse picker (bandwidth of \SI{7}{\GHz}) will be described
in more detail in the next subsection.
Furthermore, we use a second spectral filter to compensate amplifier-induced
frequency shifts \cite{Agrawal1989} of up to \SI{4}{\nm} and further limit the
bandwidth to \SI{6.4}{\nm}.
Next, a high power EDFA amplifies the
pulse train from \SI{15}{\mW} to a maximum average power of \SI{2.8}{\W}.
A free-space third order dispersion compressor reduces the pulse width to
\SI{680}{\fs} (time bandwidth product 0.53) which is close to the transform limited pulse width
of \SI{560}{\fs} for a Gaussian-shaped pulse of
the given bandwidth.

\subsection{Pulse picking}
  \label{sec:picking}

In order to select the pulse sequences described in section \ref{sec:intro}
out of the \SI{5}{\GHz} pulse train, we need an optical element that is able to
select pulses at this rate and to withstand up to \SI{2.8}{\W} of laser power after
the high power EDFA.
To satisfy both requirements, we chose a twofold approach using a fast switching
element before the high
power EDFA (where the average laser power can be limited to \SI{10}{\mW})
and a slow element after the amplifier to create the desired pulse sequences
(see panel C in figure \ref{fig:setup}).

The fast element is a pulse picker (custom-made ModBox by Photline) which contains
a Mach-Zehnder interferometer with an electro-optic modulator of
\SI{7}{\GHz} bandwidth.
Since its maximum optical input power is on the order of a few \si{\mW}, we install
it before the high power EDFA where the light intensity is sufficiently low.
Considering the amplifier's need to be seeded continuously with a maximally
allowed dark time on the order of \SI{20}{\ns},
we need an additional switching element after the amplifier with a high damage threshold
and a switching time of less than \SI{20}{\ns}.
For this we use a Pockels cell (Leysop BBO-3-25-AR790) with a driver
(custom-made by Bergmann Me\ss{}ger\"ate Entwicklung) that enables switching of the
cell with a rise/fall time of \SI{7}{\ns}
at a maximum repetition rate of \SI{10}{\MHz} and a measured optical extinction
ratio of \SI{30}{\dB}.
Both the pulse picker and the Pockels cell are controlled by an arbitrary
waveform generator (Tektronix AWG 70002A) with a sample rate of \SI{25}{GS\per\s}
which is synchronized with the seed laser by a \SI{250}{\MHz} RF signal derived
directly from the laser's pulse train.

\begin{figure}
\centering
	\includegraphics[width=0.8\textwidth]{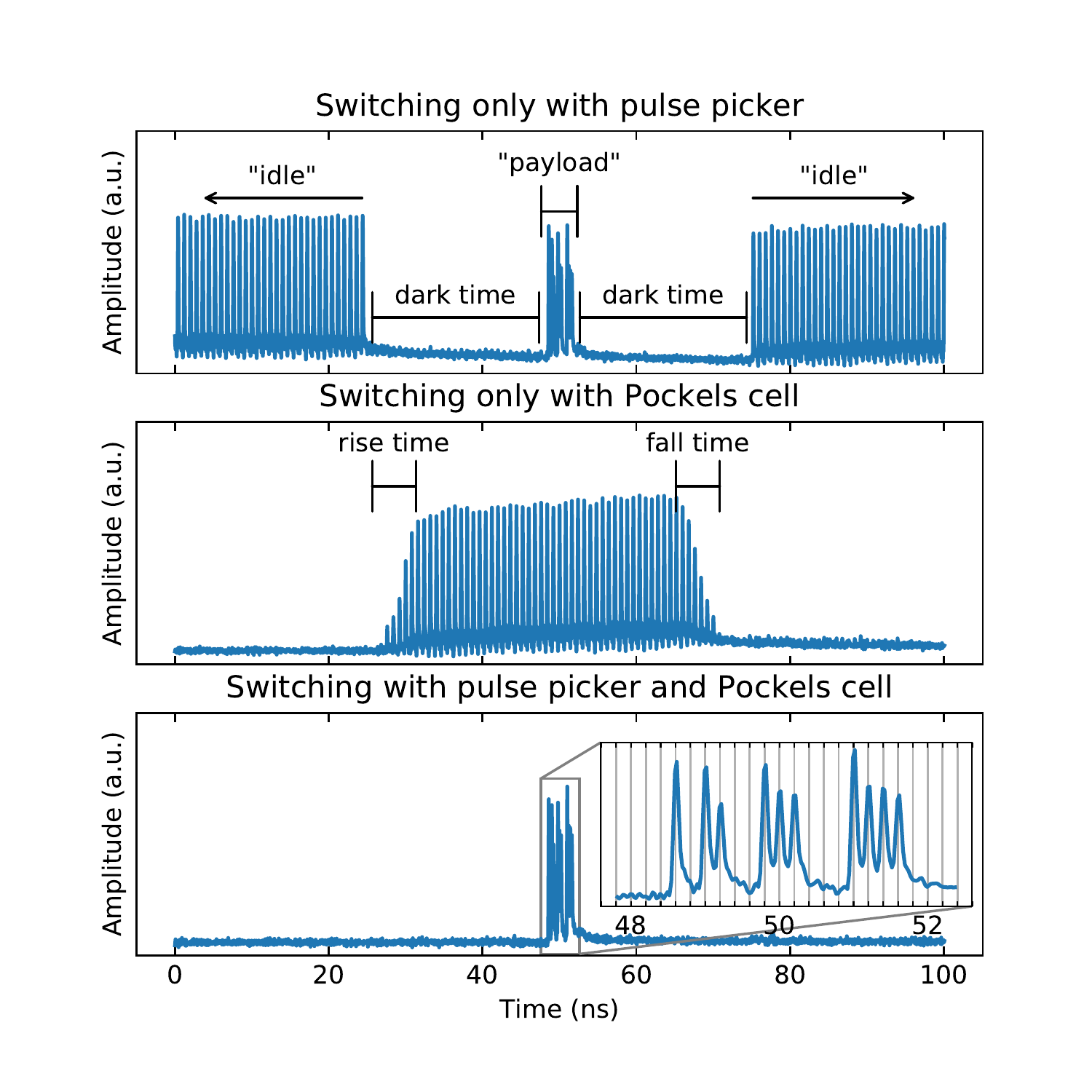}
	\caption[Pulse patterns]
	{
		Pulse patterns generated by using either only the pulse picker (top),
		only the Pockels cell (middle) or both (bottom), measured by detecting
		the residual \SI{786}{\nm} light after the PPKTP crystal.
		The pulses labeled ``idle'' on both sides of the top panel are required to
		seed the high-power EDFA.
		The length of the dark time on either side of the ``payload'' is
		determined by two factors:
		The rise and fall times of the Pockels cell of \SI{7}{\ns} and the minimum
		allowed time of \SI{35}{\ns} between switching the cell on and off
		--~i.e. between the start of the rise time and the start of the fall time.
		The inset in the bottom panel shows the zoomed-in payload signal.
		Every grid line corresponds to the location of a pulse in the original
		\SI{5}{\GHz} pulse train.
		For the reason for the different pulse heights see section
		\ref{sec:pulseswitching} and especially figure 
		\ref{fig:switch-on}.
	}
	\label{fig:pulse_picking}
\end{figure}

Figure \ref{fig:pulse_picking} shows three oscilloscope traces recording an
arbitrary pulse pattern (including a dark time) which was detected by a photodiode.
To generate the three traces, either only the pulse picker, only the Pockels cell or
both were used.
It shows that the pulse picker can switch individual pulses in arbitrary
sequences.

\subsection{Frequency up conversion}
  \label{sec:shg}

\begin{figure*}
\centering
 \includegraphics{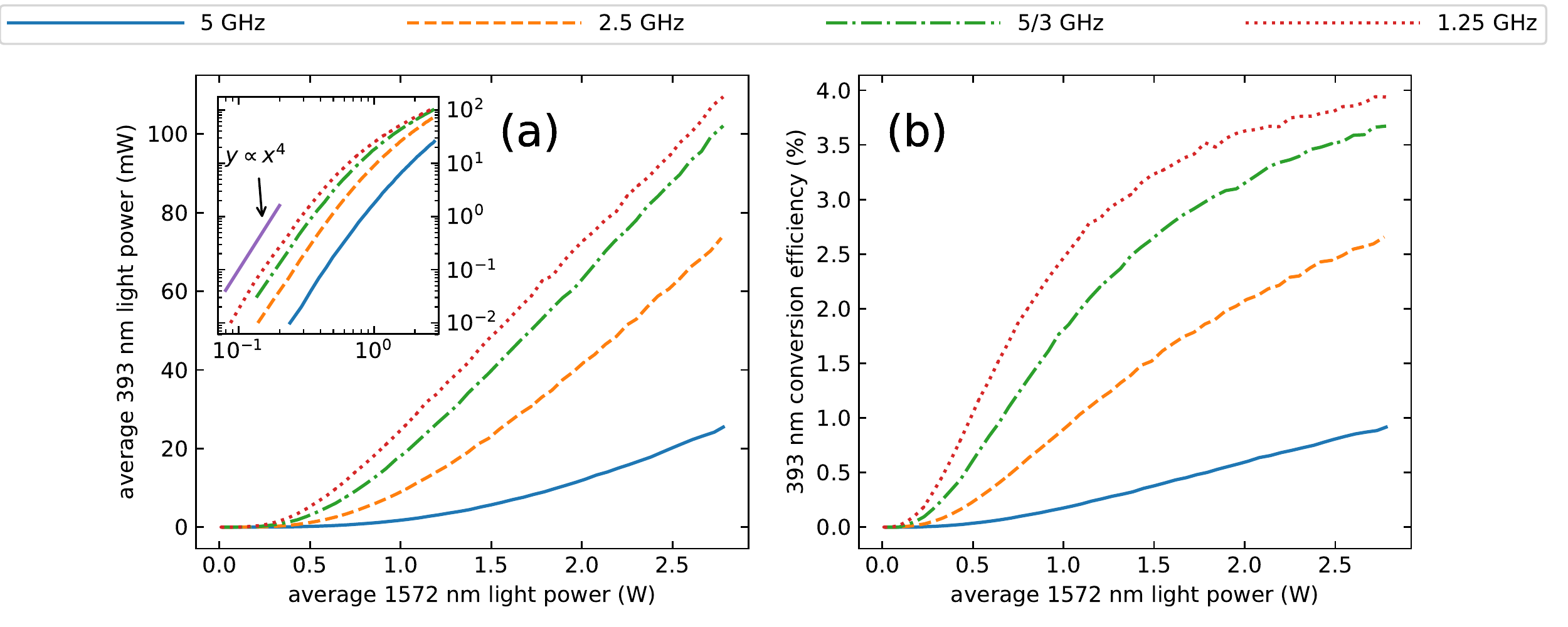}
 \caption{Measurement of \SI{393}{\nm} light power and conversion efficiencies
 of four different repetition rates
 as a function of fundamental \SI{1572}{\nm} light power.
 }
 \label{fig:conversion_efficiency}
\end{figure*}

The fundamental \SI{1572}{\nm} light is converted to \SI{393}{\nm} by frequency
doubling the light twice in two separate nonlinear crystals
(panel D in figure \ref{fig:setup}).
The first one is an MgO-doped PPLN (periodically-poled lithium niobate) crystal
(Covesion MSHG1550-0.5-5),
the second crystal is a PPKTP (periodically-poled potassium titanyl phosphate)
crystal (custom-made by Raicol Crystals Ltd.).
Since the doubling efficiency of the crystals scales with the square of the peak power,
the measured extinction ratio of the Pockels cell/PPKTP crystal system is \SI{56}{\dB},
i.e. almost the square of the Pockels cell extinction ratio.
At the current maximum average \SI{1572}{\nm} light power of about \SI{2.8}{\W}
and at a repetition rate of \SI{5}{\GHz} (\SI{1.25}{\GHz}, and therefore four
times higher peak power) we can produce
\SI{0.77}{\W} (\SI{1.18}{\W}) of \SI{786}{\nm} light, which corresponds to a
conversion efficiency of \SI{28}{\%} (\SI{42}{\%}),
and \SI{25}{\mW} (\SI{110}{\mW}) of \SI{393}{\nm} light, which corresponds to a
conversion efficiency of the second frequency doubling step of \SI{3.2}{\%}
(\SI{9.3}{\%}).
The maximum total conversion efficiency (from \SI{1572}{\nm} to \SI{393}{\nm})
is \SI{0.9}{\%} (\SI{3.9}{\%}).
Measurements of the \SI{393}{\nm} light power and conversion efficiencies for four different
repetition rates are presented in figure \ref{fig:conversion_efficiency}.
The inset in panel (a) of that figure is a log-log graph of the
same data.
The graph, together with the line $y \propto x^4$ visualizes the expected fourth-power
dependency of the \SI{393}{\nm} light power on the \SI{1572}{\nm} light power.

After each up-conversion of the laser pulses, the remaining fundamental light is
split off by a dichroic mirror and in case of the 2nd SHG stage further
suppressed by a shortpass filter (Thorlabs FESH0700).
Next, the light is coupled into a polarization-maintaining single-mode fiber
and sent to the ion trap.
After the \SI{5}{\m} long fiber, we measure a pulse width of \SI{3}{\ps}.
Finally, a collimator equipped with a \SI{250}{\mm} lens focuses the laser beam
to a waist measured to be $w_0 = \SI{11.8+-0.3}{\micro\m}$ and
directs the beam such that the ion is located in the beam waist.

Unless stated otherwise, all following measurements and experiments presented
in this paper were conducted at a high power EDFA output power of
\SI{\approx 600}{\mW}.
During times when the Pockels cell was off and the pulses directed into
a beam dump, the high power EDFA was
seeded with pulses at \SI{1.25}{\GHz} repetition rate.

\subsection{Pulse switching characteristics}
  \label{sec:pulseswitching}

\begin{figure*}
\centering
 \includegraphics{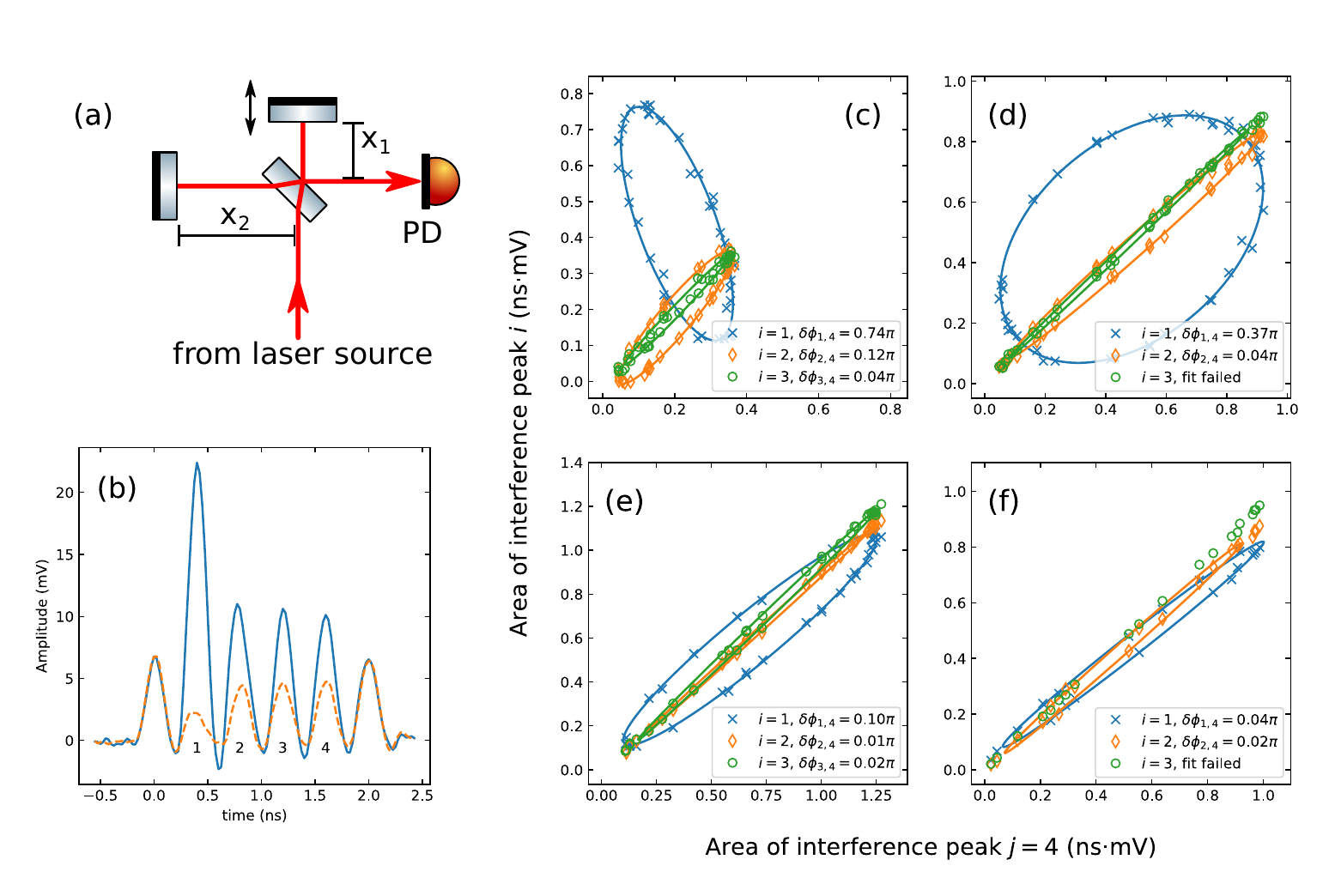}
 \caption{
  (a) Michelson interferometer for a measurement of pulse-to-pulse phase shifts.
  (b) Example signals at $f_\text{rep} = \SI{2.5}{\GHz}$.
  The labeled peaks between \SI{0.4}{\ns} and \SI{1.6}{\ns} (four of each
  plot line) are interference pulses of the first four input-pulses
  each with its successor.
  (c-f): Areas of the first (blue crosses),
  second (yellow diamonds) and third (green circles) interference peak
  plotted against the area of the fourth interference peak.
  The solid curves are a fit of the ellipse
  $\left\lgroup x(\Delta \phi) = c * \sin(\Delta \phi) + x_0,
  \;y(\Delta \phi) = c * \sin(\Delta \phi + \delta \phi_{i,j}) + y_0\right\rgroup$
  to the data, which allows us to extract $\delta \phi_{i,j}$
  up to a sign, since for $\delta \phi_{i,j} = \Phi$ and
  $\delta \phi_{i,j} = - \Phi$ the ellipses are congruent.
  The repetition rate is \SI{5}{\GHz} (c), \SI{2.5}{\GHz} (d), \SI{5/3}{\GHz} (e)
  and \SI{1.25}{\GHz} (f), respectively.
  For higher repetition rates, the phase difference between the first and second
  pulses is larger than the phase difference between later pulses.
  At lower repetition rates, the phase difference between any two consecutive
  pulses is equal within \SI{1}{\percent} of $2\pi$.
 }
 \label{fig:interferometer_and_ellipses}
\end{figure*}

Our pulse picking scheme described in subsection \ref{sec:picking} necessitates
blocking the pulses with the fast pulse picker during the rise and fall time
of the Pockels cell for typically \SI{12}{\ns}.
This results in an equally long dark time of the amplifiers after the pulse
picker.
We examined the characteristics of the pulses during the first nanoseconds
after such dark time by measuring pulse areas and relative phase of the pulses.

We characterize the phase shifts of the pulses by interfering them 
in a Michelson interferometer
(see figure \ref{fig:interferometer_and_ellipses}(a)).
At the beam splitter the input pulse train is split into two copies,
one of which is temporally delayed by one pulse period $\tau_\text{pulse}$
with respect to the other before being recombined.
In this way, every pulse $i$ interferes with its successive pulse $i+1$ and
the total phase difference $\Delta \Phi_i$ of the two pulses at the output of
the interferometer is a function of the phase difference
$\Delta \phi_i$ of pulses $i$ and $i+1$,
and the path length difference
$\Delta x = 2x_2 - 2x_1 \approx c \cdot \tau_\text{pulse}$:
\[
 \Delta \Phi_i = \Delta \phi_i + k \cdot \Delta x,
\]
with $x_1$ ($x_2$) the length of interferometer arm 1 (2),
$c$ the speed of light and $k$ the length of the wave vector.
The interference pulses are detected on a fast photodiode and their pulse areas,
which are functions of $\Delta \Phi_i$, extracted by integrating the
photodiode's signal over the pulse length.

We can tune $\Delta x$ by manually moving one of the retro-reflecting mirrors
which is fixed on a manual translation stage,
but we can not deterministically change $\Delta x$ on a sub-wavelength scale
which would have allowed to keep track of the changes in $\Delta \Phi_i$ due to
changes in $\Delta x$.
Nevertheless, we found its fluctuations to be much smaller than the light's
wavelength and consider $\Delta x$ to be constant on time scales of our
measurements of about \SI{100}{\ms}.
For each measurement we therefore randomly choose $\Delta x$ and repeat the
measurement for different $\Delta x$.
Accordingly, the interference pulse areas of each measurement are both random
but also correlated in size since $\Delta x$ is random but equal for all
pulses in a given measurement, and
if the pulse areas of two interference pulses $i$ and $j$ are different,
there is also a change $\delta \phi_{i,j} = \Delta \phi_i - \Delta \phi_j$
in their phase differences.

In order to determine $\delta \phi_{i,j}$ we repeatedly send
a five-pulse pulse train into the interferometer at a rate of \SI{1}{\kHz} and
average over 100 consecutive trains.
We measure the interference pulse areas for different and random $\Delta x$
by moving the translation stage back and forth on the order of
\SI{5}{\micro\m} between two 100 pulse trains-long measurements.
Figure \ref{fig:interferometer_and_ellipses}(b) shows two example interference
signals, taken at \SI{2.5}{\GHz}.
For every measurement at different $\Delta x$ we plot the pulse areas of two
interference pulses against each other and present the data in panels
(c) through (f) in figure \ref{fig:interferometer_and_ellipses}.
Every abscissa is the area of interference pulse 4;
the ordinates are the areas of interference pulses
1 through 3, respectively.

The data points fall on a straight line at a \ang{45} angle,
if $\delta \phi_{i,j}$ between the two interference pulses is zero or an
integer multiple of $2\pi$, and at a \ang{-45} angle if
$\delta \phi_{i,j} = (2n+1) \pi, n \in \mathbb{Z}$.
In general, for any other value of $\delta \phi_{i,j}$
the data points are located on an ellipse.
This allows us to extract $\delta \phi_{i,j}$ by fitting an ellipse to
the data.

We see a dependence of $\delta \phi_{i,j}$ on the length of the pulse period
$\tau_\text{pulse} = 1 / f_\text{rep}$ if there was a long
(here: \SI{12}{\ns}) dark time before the first pulse:
For the maximum repetition rate of $f_\text{rep} = \SI{5}{\GHz}$ we find that
between the first (second) and fourth interference peak
(pulses 1\&2 (2\&3) and 4\&5, respectively)
$\delta \phi_{1,4} = \SI{0.74}{\pi}$ ($\delta \phi_{2,4} = \SI{0.12}{\pi}$).
For later interference peaks ($i \geq 3$) $\delta \phi_{i,4}$ is vanishingly small.
As a trend we observe that for larger $\tau_\text{pulse}$, $\delta \phi_{i,j}$
becomes smaller and is vanishing for
$\tau_\text{pulse} \geq \SI{800}{\ps} = 1 / \SI{1.25}{\GHz}$.

\begin{figure}
\centering
 \includegraphics{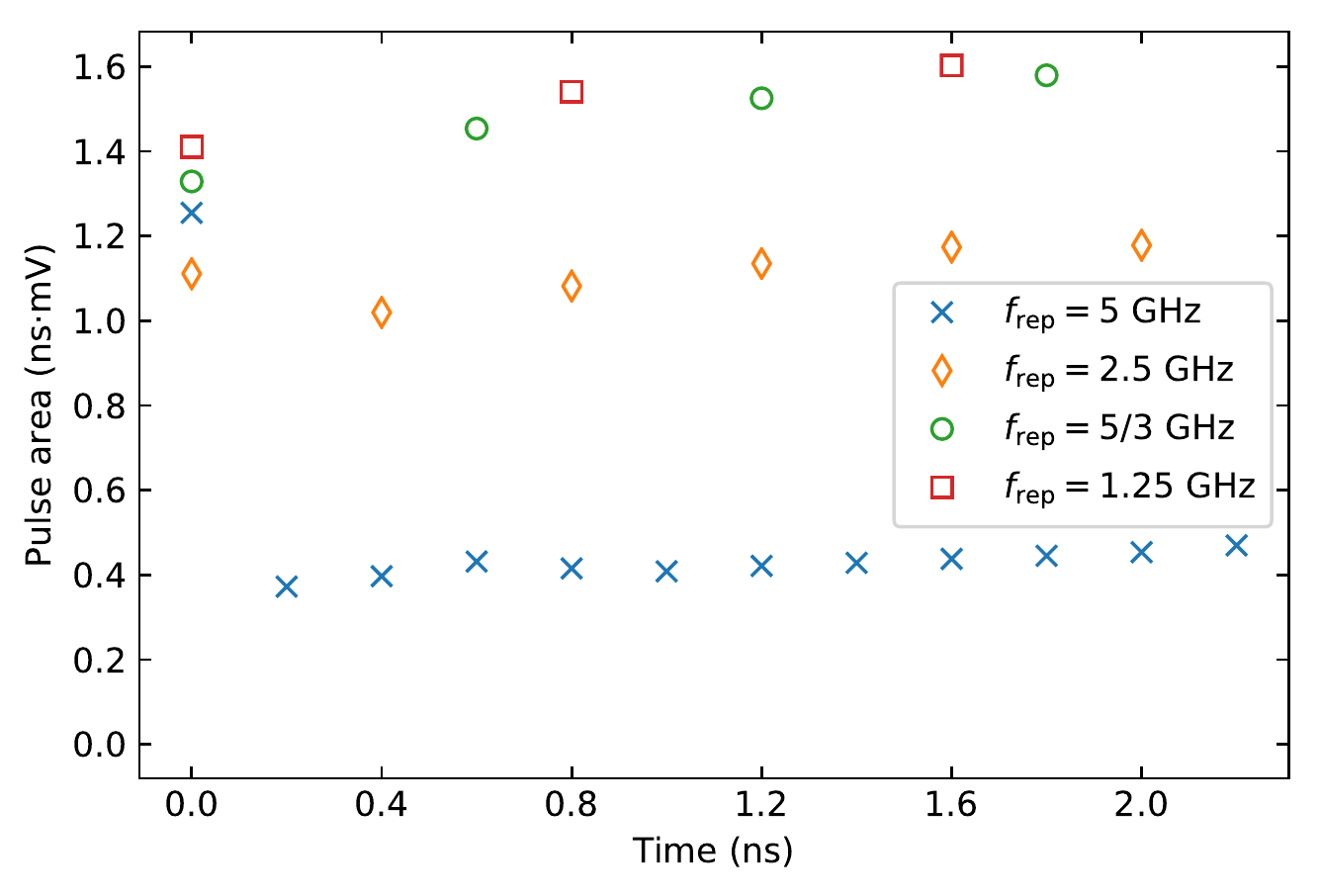}
 \caption{
  Pulse areas of pulses during the first \SI{2}{\ns} after a \SI{12}{\ns}
  dark time at repetition rates from \SIrange{1.25}{5}{\GHz}.
  Time axis relative to first pulse.
  Data points are the average of \num{1000} measurements, error bars are
  smaller than data point symbols.
  Average high power EDFA output power for this measurement was
  \SI{\approx 1.6}{\W}.
 }
 \label{fig:switch-on}
\end{figure}

In addition to the phase shift, we observe repetition rate-dependent
deviations of the pulse intensity of the first pulses after the dark time.
Figure \ref{fig:switch-on} shows the pulse areas of the pulses during
the first \SI{2}{\ns} after the dark time.
In the case of \SI{5}{\GHz} repetition rate we observe
that all pulses from the second onward are weaker by a factor of about 3 with
respect to the first pulse.
At \SI{2.5}{\GHz} repetition rate, the intensity of the second pulse decreases
by about \SI{10}{\%} with respect to the first.
At the other repetition rates we do not observe this effect.
For all repetition rates we observe an increase in pulse intensity
of about \SI{10}{\%}
for the shown time of \SI{2}{\ns}
of pulses after the possible initial decrease.

Both the phase shift and the change in pulse area will be problematic for
experiments using coherent manipulations of the qubit.
Since for our planned two-ion phase gate every pulse needs to act as a
$\pi$-pulse, the pulse area needs to be constant, but due to the way we plan
to create the counter-propagating pulse pairs, the phase shift does not:
We intent to generate the two pulses of every required $\pi$-pulse pair
by splitting a larger-area pulse.
Therefore, every pulse pair will consist of identical copies and the two pulses
will add up to a $2\pi$-rotation of the Bloch vector around the same axis
regardless of phase difference with respect to other pulse pairs.
For this reason only the intensity changes
should pose an issue for our phase gate, the phase shifts should not.

We believe that both effects are caused by a semiconductor optical amplifier
(SOA2 in figure \ref{fig:setup}) which serves as a preamplifier to the
high power EDFA and are due to the finite carrier lifetime of the SOA of
\SI{500}{\ps}.
This causes the dynamical behavior of the SOA to depend on the input signal 
of the past \SI{\sim 500}{\ps} \cite{Manning1997} and is known as the
``pattern effect''\cite{Rizou2016}.
We therefore plan to replace the SOA with another fiber-based preamplifier.

\section{Methods to measure the Rabi frequency of a pulsed laser-ion system}
  \label{sec:theory}

We have employed three different techniques to measure the rotation angle
per pulse $\theta = \Omega \, \delta t$.
The model for the ion-laser interaction used for simulating our experiments
is presented in subsection \ref{sec:model}.
Subsection \ref{sec:results} presents measurements using the three techniques.
The first approach uses many pulses -- $\order{\num{e3}}$ --
to pump the ion into the metastable \Dfivehalf{} state.
Measuring the population in that state and fitting the model to the data
allows $\theta$ to be extracted.
Next, we use the same principle but in the regime of single pulses,
allowing pulse dynamics and characteristics to be extracted.
Finally, the third method is based on preparing the ion in a coherent
superposition of the two qubit states.
We then use single laser pulses to reduce and restore
coherence between the qubit states by transferring population of one of the two
states back and forth to a third state.

\subsection{Pulse model for simulation and fits}
  \label{sec:model}

We calculate the interaction of a \Ca{} ion with a pulsed laser field
resonant to the ion's \trans{\Sonehalf}{\Pthreehalf} transition
(see figure \ref{fig:levelscheme}).
We consider spontaneous decay of the \Pthreehalf{} state
into the \Sonehalf{} state with decay rate $\Gamma_\text{PS}$ and
into the \Dfivehalf{} state with decay rate $\Gamma_\text{PD}$.
For the experiments that we describe, the \Dthreehalf{} state does not need
to be considered as any population in this state is pumped back
to \Sonehalf{} via \Ponehalf{} on a timescale of \SI{10}{\ns}
with an \SI{866}{\nm} laser. 
Therefore, any decay to \Dthreehalf{} effectively becomes a decay to
\Sonehalf{} and the modified decay rates are
\begin{equation}
 \Gamma_\text{PS} = (1 - p_{5/2}) \, \Gamma,
\end{equation}
\begin{equation}
 \Gamma_\text{PD} = p_{5/2} \, \Gamma,
\end{equation}
where $\Gamma = \frac{1}{\tau} = \frac{1}{\SI{6.924+-0.0019}{\ns}}$ \cite{Jin1993} and
$p_{5/2} = \num{0.0587+-0.0002}$ the probability of \Pthreehalf{} to decay to \Dfivehalf{}
\cite{Gerritsma2008}.

Consequently, we assume a three-level system and identify
the \Sonehalf{} state with $\ket{1}$,
the \Pthreehalf{} state with $\ket{2}$ and
the \Dfivehalf{} state with $\ket{3}$,
as indicated in figure \ref{fig:levelscheme}.
Furthermore, we assume infinitesimally short pulses
and neglect any decay of $\ket{3}$, since the \Dfivehalf{} state's lifetime
of \SI{1.2}{\s} \cite{Kreuter2005} is much longer than the cycle time of our
experiments of about \SI{10}{\ms}.
The system's excited state $\ket{2}$ decays spontaneously with rates
$\Gamma_{21} \equiv \Gamma_\text{PS}$ and $\Gamma_{23} \equiv \Gamma_\text{PD}$
into state $\ket{1}$ and $\ket{3}$, respectively.

To model the time evolution of the quantum state over
one pulse period of duration $\tau_\text{pulse} = 1 / f_\text{rep}$
we start with a rotation of the Bloch vector of the
$\ket{1}$ - $\ket{2}$ subsystem around the x-axis to account for the effect of
a single pulse, followed by a rotation around the z-axis to account for
the detuning during the dark time between two pulses.
Next, a rotation of the Bloch vector of the $\ket{1}$ - $\ket{3}$ subsystem
around its z-axis takes into consideration an effect of the pulses on this
transition's frequency.
Finally, we account for spontaneous decay of state $\ket{2}$ during the pulse
period by applying the appropriate Kraus operators to the quantum state.

Hence, we start by applying an x-rotation operator $\mathcal{U_R}$ to a
given density operator $\rho_0$:
\begin{equation}
 \rho' = \mathcal{U_R} \rho_0 \mathcal{U_R}^\dag,
\end{equation}
with
\begin{equation}
 \mathcal{U_R} =  \exp\left(\frac{\mathrm{i}}{2} \ \theta \ (\ket{2}\bra{1} + \ket{1}\bra{2})\right),
 \label{eq:R}
\end{equation}
where $\theta$ is the rotation angle.

Next, we allow for a possible detuning $\Delta$ of the laser light with respect
to the \trans{1}{2} transition and a shift $\Delta^\prime$ of the state $\ket{3}$
by applying a z-rotation $\mathcal{U_Z}$ to the result.
Here, the angle of rotation is proportional to $\tau_\text{pulse}$, $\Delta$
and $\Delta^\prime$, respectively:
\begin{equation}
 \rho'' = \mathcal{U_Z} \rho' \mathcal{U_Z}^\dag,
\end{equation}
with
\begin{equation}
 \mathcal{U_Z} =  \exp(\mathcal{Z}),
\end{equation}
\begin{equation}
 \mathcal{Z} = \frac{\mathrm{i}}{2} \left[\Delta \ (\ket{1}\bra{1} - \ket{2}\bra{2}) +
 \Delta^\prime \ (\ket{1}\bra{1} - \ket{3}\bra{3})\right] \tau_\text{pulse}.
\end{equation}

Finally, we calculate the Kraus operators $\mathcal{N}$, $\mathcal{D}$ to allow
for spontaneous decay of the excited state $\ket{2}$:
\begin{equation}
 \mathcal{N} = \ket{1}\bra{1} + \sqrt{1 - p - q} \, \ket{2}\bra{2} + \ket{3}\bra{3},
\end{equation}
\begin{equation}
 \mathcal{D} = \sqrt{p} \, \ket{1}\bra{2} + \sqrt{q} \, \ket{3}\bra{2},
\end{equation}
\begin{equation}
 \rho''' = \mathcal{N} \rho'' \mathcal{N}^\dag + \mathcal{D} \rho'' \mathcal{D}^\dag,
\end{equation}
with
\begin{equation}
 p = 1 - \exp\left(-\Gamma_{21} \, \tau_\text{pulse}\right),
\end{equation}
\begin{equation}
 q = 1 - \exp\left(-\Gamma_{23} \, \tau_\text{pulse}\right).
\end{equation}
Calculating the decay of the excited state only after applying the x- and
z-rotation operators is acceptable, since the pulse length of \SI{3}{\ps} is very short
compared to the state's lifetime of \SI{6.9}{\ns}, and the z-rotation and decay do not
influence each other.

To find the density operator $\rho_n$ after a train of $n$ pulses we
iteratively apply these operators $n$ times
\begin{equation}
 \rho_n =
   \mathcal{N} \, \mathcal{U_Z} \, \mathcal{U_R} \, \rho_{n-1}
   \, \mathcal{U_R}^\dag \, \mathcal{U_Z}^\dag \, \mathcal{N}^\dag  +
   \mathcal{D} \, \mathcal{U_Z} \, \mathcal{U_R} \, \rho_{n-1}
   \, \mathcal{U_R}^\dag \, \mathcal{U_Z}^\dag \, \mathcal{D}^\dag
 \label{eq:simulation}
\end{equation}
and finish the calculation by letting state $\ket{2}$ decay completely.
From the resulting density operator we can easily calculate experimentally
accessible observables such as populations $\Tr(\ket{1}\bra{1}\rho_n)$, $\Tr(\ket{3}\bra{3}\rho_n)$
and coherences $\Tr(\ket{1}\bra{3}\rho_n)$, $\Tr(\ket{3}\bra{1}\rho_n)$.

To account for our observations described in section \ref{sec:pulseswitching} we
allow the first pulse in a pulse train to have a different (usually higher)
peak power and therefore allow for its
rotation angle $\theta^\text{1st}$ to be different.
Additionally, we allow the first pulse to turn the Bloch vector of the
subsystem around an axis in the equatorial plane rotated by $\delta \phi_{1,4}$ from the
x-axis to account for its phase offset relative to the other pulses.
We therefore replace $\mathcal{R}$ with $\mathcal{R}^\text{1st}$ in equation
\ref{eq:R}:
\begin{equation}
 \mathcal{R} \rightarrow \mathcal{R}^\text{1st} =
 \frac{\mathrm{i}}{2} \, \theta^\text{1st}
 \left(\cos(\delta \phi_{1,4}) \sigma_x^{12} + \sin(\delta \phi_{1,4}) \sigma_y^{12}\right).
\end{equation}

\subsection{Experimental results}
  \label{sec:results}

Experiments are conducted in the same linear Paul trap as described in
\cite{Guggemos2015}.
Its trap axis is aligned with the quantization axis defined by a bias magnetic
field.
Circularly polarized \SI{393}{\nm} laser pulses that are sent through the holes in
the trap's endcap electrodes therefore couple only pairs of Zeeman states of the
ground and excited state.

\subsubsection{Pumping into a dark state with many pulses.}
  \label{sec:rabi}

We send pulse trains of between 500 and 5000 pulses to the Doppler-cooled
and optically pumped ion
and collect fluorescence photons on the \trans{\Sonehalf}{\Ponehalf}
transition in order to determine whether the ion has decayed into the
dark \Dfivehalf{} state.
We repeat the measurement 100 times to statistically determine the
probability that the ion has decayed into the \Dfivehalf{} state.
Due to the large number of pulses, the pulse switching effects described in section
\ref{sec:pulseswitching} and which are affecting only the first one or two pulses,
can be disregarded.
For the same reason the experiment does not necessitate the use
of a pulse picker that is faster than the fundamental repetition rate.

\begin{figure}
\centering
 \includegraphics{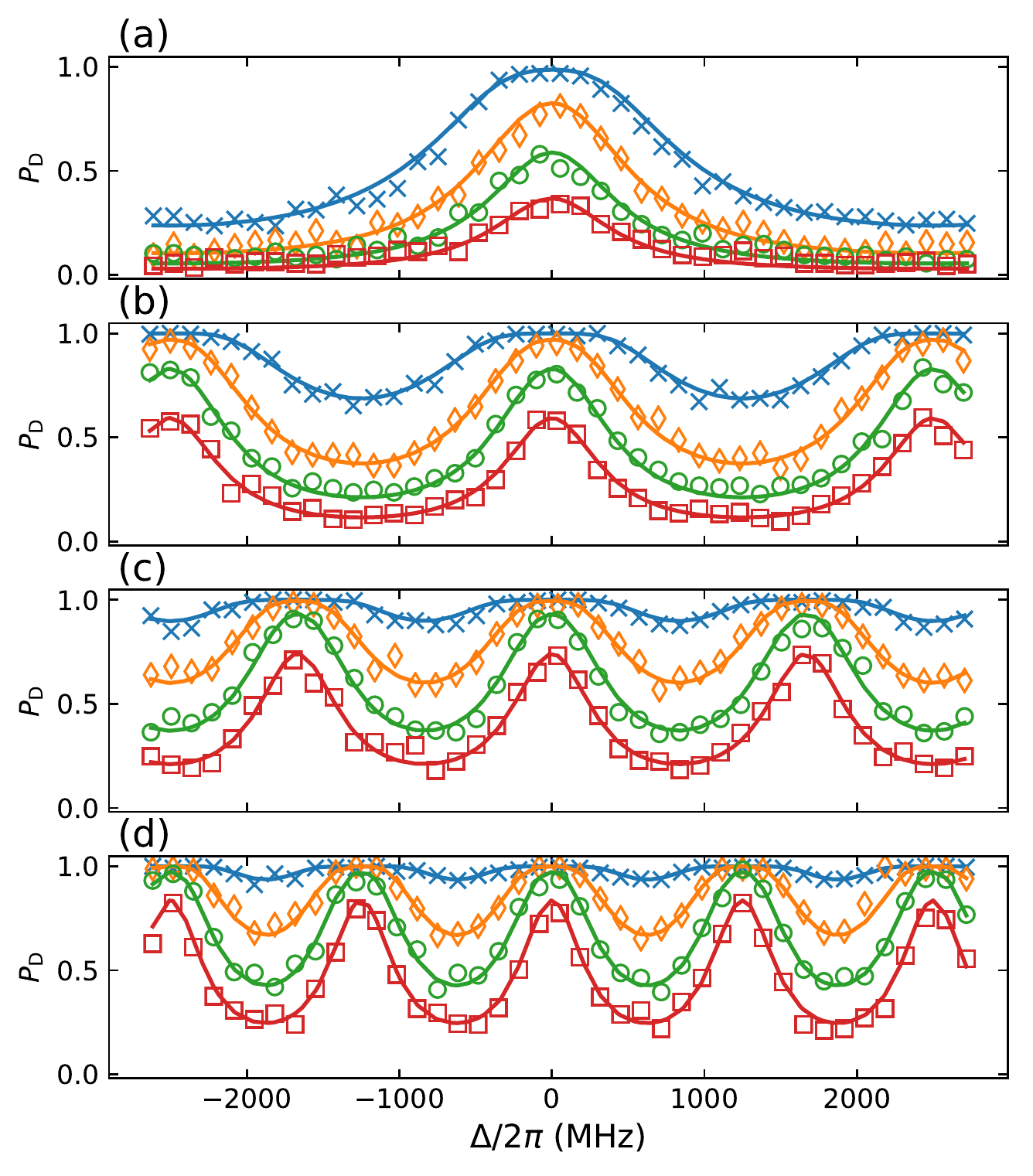}
 \caption{
  Probability $P_\text{D}$ to find the ion in the \Dfivehalf{} state as a function of the
  detuning $\Delta$ of the laser frequency with respect to the
  \trans{\Sonehalf}{\Pthreehalf} transition frequency.
  Blue crosses using \num{5000} pulses,
  yellow diamonds \num{2000} pulses,
  green circles \num{1000} pulses,
  red squares \num{500} pulses.
  The lines in each graph are a model fit to the respective data
  as described in the text,
  allowing the determination of $\theta$
  (a) \SI{5}{\GHz} repetition rate, for which we find
  $\theta = \SI{0.227+-0.012}{\pi}$.
  (b) \SI{2.5}{\GHz} repetition rate,
  $\theta = \SI{0.323+-0.005}{\pi}$.
  (c) \SI{5/3}{\GHz} repetition rate,
  $\theta = \SI{0.363+-0.005}{\pi}$.
  (d) \SI{1.25}{\GHz} repetition rate,
  $\theta = \SI{0.345+-0.005}{\pi}$.
 }
 \label{fig:rabi}
\end{figure}

We measure the \Dfivehalf{} state population $P_\text{D}$
as a function of the laser detuning
for different pulse train lengths.
Figure \ref{fig:rabi} shows data sets for
repetition rates ranging from \SIrange{1.25}{5}{\GHz}.
Each data set consists of four data sub-sets which differ only in the number
of pulses.
In order to extract $\theta$ we fit the simulation in equation
(\ref{eq:simulation}) simultaneously to each of the four sub-sets
such that we get a single value for $\theta$.
The only other fit parameter is a detuning offset which is eliminated in the
figure.
For the repetition rate of
\SI{5}{\GHz} (\SI{2.5}{\GHz}, \SI{5/3}{\GHz}, \SI{1.25}{\GHz})
we obtain $\theta = \SI{0.227+-0.012}{\pi}$
(\SI{0.323+-0.005}{\pi}, \SI{0.363+-0.005}{\pi}, \SI{0.345+-0.005}{\pi}).

\subsubsection{Pumping into a dark state with single pulses.}
  \label{sec:singlepulses}

\begin{figure}
\centering
 \includegraphics{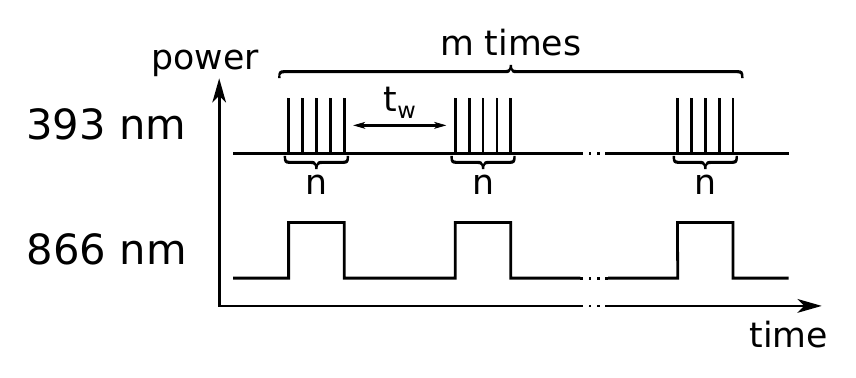}
 \caption{
  Experimental sequence of dark state pumping with single pulses.
  The ion is Doppler-cooled and prepared in the \Sonehalf{} state.
  A pulse train of $n$ pulses coherently drives the
  \trans{\Sonehalf}{\Pthreehalf} transition.
  The pulse train is repeated a total of $m=20$ times with a waiting time $t_\text{w}$
  between each two repetitions.
  We choose $t_\text{w} = \SI{20}{\micro\s} \gg \tau = \SI{6.924}{\ns}$
  with $\tau$ the \Pthreehalf{} state's lifetime.
  Finally, we measure the \Dfivehalf{} state probability.
 }
 \label{fig:single_pulses_sequence}
\end{figure}

\begin{figure}
\centering
 \includegraphics{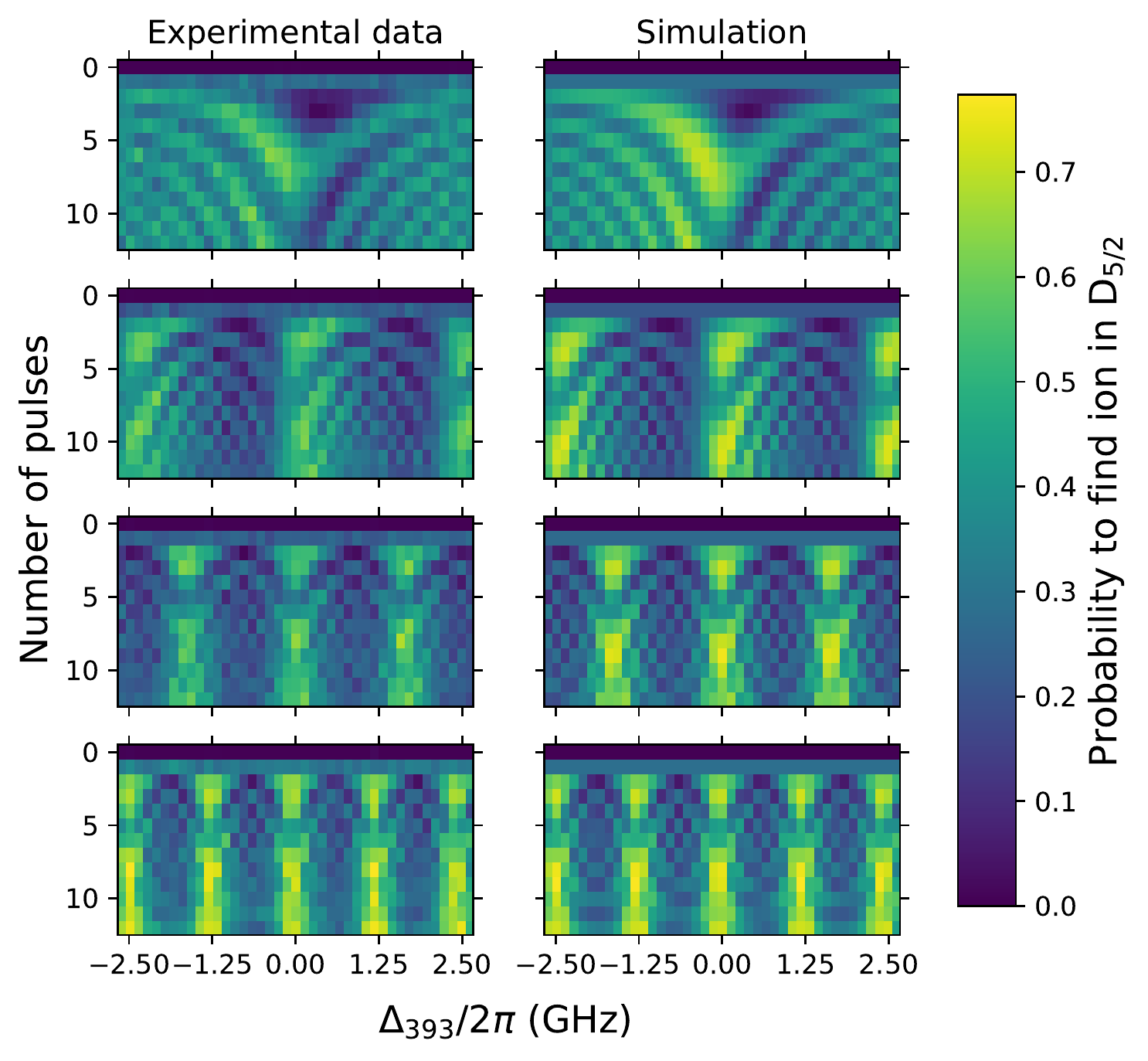}
 \caption{
  Probability to excite the ion to \Dfivehalf{} with different number of pulses
  and detuning. Left column: experimental data; right column: simulation/fit.
  1st row: \SI{5}{\GHz} repetition rate,
  $\theta^\text{1st} = \SI{0.353+-0.001}{\pi}$,
  $\theta = \SI{0.195+-0.018}{\pi}$, $\delta \phi_{1,4}=\SI{1.282+-0.001}{\pi}$.
  2nd row: \SI{2.5}{\GHz} repetition rate,
  $\theta = \SI{0.312+-0.010}{\pi}$, $\delta \phi_{1,4}=\SI{0.361+-0.002}{\pi}$.
  3rd row: \SI{5/3}{\GHz} repetition rate,
  $\theta = \SI{0.339+-0.007}{\pi}$, $\delta \phi_{1,4}=\SI{0.088+-0.003}{\pi}$.
  4th row: \SI{1.25}{\GHz} repetition rate,
  $\theta = \SI{0.358+-0.003}{\pi}$, $\delta \phi_{1,4}=\SI{0.051+-0.001}{\pi}$.
  The asymmetry evident in the first two rows is due to the phase shift of
  the first pulse with respect to later pulses.
 }
 \label{fig:singlepulses}
\end{figure}

In order to implement our phase gate scheme we need to ensure that every
single pulse is a $\pi$-pulse and it does not suffice to characterize an
ensemble of hundreds of pulses.
By instead sending only $n \leq 12$ pulses at a time to the ion
we can gain crucial insights into single pulse dynamics and characteristics.
In order to prevent the problem of having to measure very small \Dfivehalf{}
state populations, we amplify the signal by repeating the pulse train $m = 20$ times
as shown in the experimental sequence in figure \ref{fig:single_pulses_sequence}.
Between repetitions, a waiting time $t_\text{w} = \SI{20}{\micro\s}$ much larger
than the \Pthreehalf{} state's lifetime $\tau = \SI{6.924}{\ns}$ ensures that any
population in that state has decayed
and we accumulate population in the \Dfivehalf{} state \cite{Gerritsma2008}.

Figure \ref{fig:singlepulses} shows the probability for being in the
\Dfivehalf{} state, inferred from our experimental data and fit/simulation
side-by-side.
We vary the detuning $\Delta$ and the number of pulses for different
repetition rates.
Free fit parameters are $\theta$, the phase offset
of the first pulse $\delta \phi_{1,4}$ and a detuning offset.
Since we know from earlier measurements described in section
\ref{sec:pulseswitching}
that the first pulse in the case of \SI{5}{\GHz} repetition rate has a
different area than the other pulses,
we allow an additional fit parameter in that case:
the rotation angle of the first pulse $\theta^\text{1st}$.
All fit values are within three standard deviations of those acquired
previously with long pulse trains (section \ref{sec:rabi}),
as well as those acquired with the Michelson interferometer (section \ref{sec:pulseswitching}).
Contrary to measuring $\delta \phi_{i,j}$ with the
interferometer, this measurement is able to also determine the sign of $\delta \phi_{i,j}$
and therefore to distinguish between phases
$\delta \phi_{i,j} = \Phi$ and $\delta \phi_{i,j} = -\Phi$ (the later being the same as
$\delta \phi_{i,j} = 2\pi - \Phi$).
We therefore assume that $\delta \phi_{1,4}$ at \SI{5}{\GHz} is about
\SI{1.25}{\pi} and not \SI{0.75}{\pi} as the ellipse fit suggested.
Furthermore, $\theta^\text{1st}$ is found to be a factor of about $\sqrt{3}$ larger
than $\theta$ for the \SI{5}{\GHz} repetition rate, which is the expected
amount for a three times larger pulse area
(also compare with section \ref{sec:pulseswitching}).

\subsubsection{Single pulse with area $\pi$.}
  \label{sec:pipulse}

In order to check if a single pulse can act as a $\pi$-pulse we repeat the
previous experiment with only a single pulse while varying the \SI{393}{\nm}
light power ($n=1$, $m=15$).
From the experimentally determined \Dfivehalf{} state probability $P_\text{D}$
after the 15 repetitions of a single pulse
we calculate the \Pthreehalf{} state probability $P_\text{P}$ after only one
single pulse using
\begin{equation}
 P_\text{P} = \frac{1}{0.0587}
 \left(1 - \left(1 - P_\text{D}\right)^\frac{1}{m}\right).
\end{equation}
Due to measurement fluctuations,
this can sometimes lead to unphysical values of $P_\text{P} > 1$ if we measure
a $P_\text{D}$ that happens to be larger than the maximum expectation value of
$P_\text{D, max} = 1-(1-0.0587)^{15} \approx 0.60$.

\begin{figure}
\centering
 \includegraphics[width=0.8\textwidth]{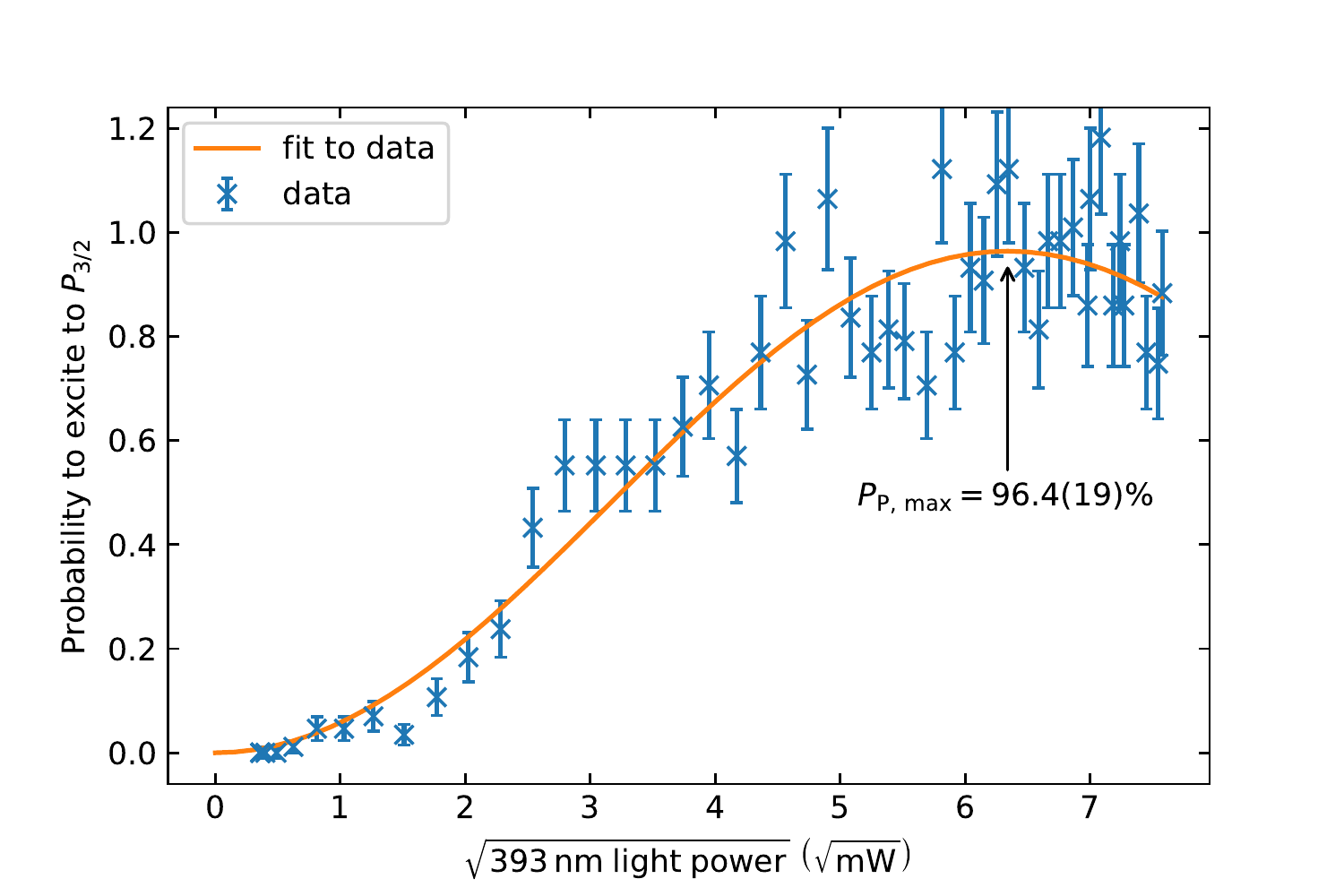}
 \caption{
  Probability to excite the ion to \Pthreehalf{} with a single pulse as a
  function of \SI{393}{\nm} light power.
  $P_\text{P, max}$ is the maximum excitation probability to the
  \Pthreehalf{} state and determined by a fit to the data.
  Measurement noise can cause unphysical values of $P_\text{P} > 1$ 
  (see text).
 }
 \label{fig:pipulse}
\end{figure}

Since the excitation probability is a function of the sine squared of the
Rabi frequency $\Omega$, we plot $P_\text{P}$ versus the square root of
the \SI{393}{\nm} light power, which is proportional to $\Omega$
(and $\theta$) in figure \ref{fig:pipulse}.
The data points should therefore follow the curve
\begin{equation}
 P_\text{P} = P_\text{P, max} \cdot \sin^2(\sqrt{P_\text{light}} / \omega),
\end{equation}
where $\omega$ is a proportionality factor of dimension \si{\sqrt{\W}} and
$P_\text{P, max}$ is the maximum probability of a single pulse exciting the ion
to the \Pthreehalf{} state.
A fit of this curve to our data yields $P_\text{P, max} = \SI{96.4+-1.9}{\%}$,
showing how close we are to achieving single pulse $\pi$-pulses.

Please note that the maximum \SI{393}{\nm} light power plotted in figure
\ref{fig:conversion_efficiency} (e) (\SI{110}{\mW}) and figure \ref{fig:pipulse}
($(\SI{7.6}{\sqrt{\mW}})^2 \approx \SI{58}{\mW}$) was measured at the same
fundamental light power.
The discrepancy is due to coupling losses in the fiber guiding the
\SI{393}{\nm} light to the ion.

\subsubsection{Ramsey contrast decay and revival.}
  \label{sec:ramsey}

In experiments using coherent manipulations of the qubit it is important to
know the effect of the manipulation on the relative phase of the two
qubit states.
The experiments of the previous three subsections used only one of the two qubit
states and the relative phase appeared only as an unobservable global phase.
Here however, the second qubit state serves as a phase reference,
giving us access to the desired information.

We let single pulses interact with the ion during the waiting time of a
Ramsey experiment
and monitor the coherence at the end of the Ramsey sequence in the following way:
We start by bringing the ion into a coherent superposition of two states by
a $\frac{\pi}{2}$-pulse on the \trans{\Sonehalf}{\Dfivehalf} transition
with a \SI{729}{\nm} laser as illustrated in figure \ref{fig:ramsey_sequence}:
\begin{figure}
\centering
 \includegraphics{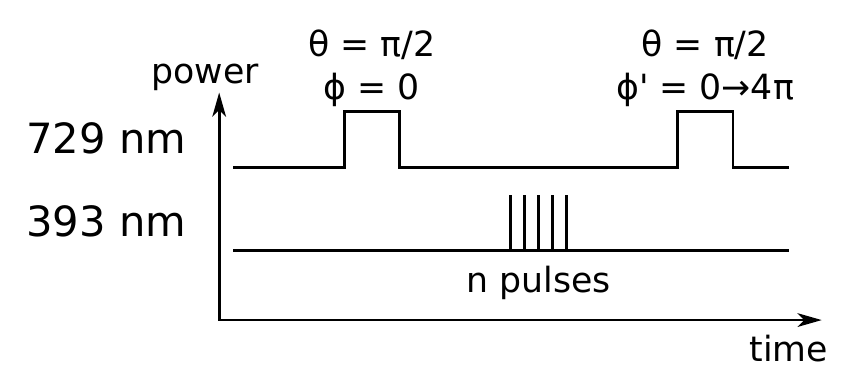}
 \caption{
  Sequence of a Ramsey contrast decay and revival experiment.
  The ion is Doppler-cooled and prepared in the \Sonehalf{} state.
  The first \SI{729}{\nm} laser pulse creates a coherent
  superposition of the ion's internal \Sonehalf{} and \Dfivehalf{} states.
  A controlled and variable number of \SI{393}{\nm} laser pulses act on the
  remaining \Sonehalf{} state population, transferring population to the
  \Pthreehalf{} state and possibly back, potentially reducing the S-D coherence.
  We finally analyze how much coherence remains by
  varying the phase of the second \SI{729}{\nm} laser pulse and
  measuring the \Dfivehalf{} state probability, while keeping the Ramsey time
  $t_R = \SI{10}{\micro\s}$
  much shorter than the coherence time of the superposition ($\order{\SI{1}{\ms}}$).
 }
 \label{fig:ramsey_sequence}
\end{figure}
\[
 \ket{\psi} = \frac{1}{\sqrt{2}} \left(\ket{1} + \ket{3}\right)
 \equiv \frac{1}{\sqrt{2}} \left(\ket{\Sonehalf} + \ket{\Dfivehalf}\right)
\]
We then use single pulses of our \SI{393}{\nm} laser system to coherently drive the
\trans{\Sonehalf}{\Pthreehalf} transition.
If there is any population remaining in the \Pthreehalf{} state after the pulses,
this part of the population will undergo spontaneous decay and thus destroy
the coherence.
We finally analyze how much coherence of the original state remains and also
extract the phase between the two states.

In mathematical terms, let $\rho$ be the density matrix describing our system
after the \SI{393}{\nm} pulses and $\sigma_x^\text{SD}$, $\sigma_y^\text{SD}$
the Pauli matrices acting only on the states
$\ket{\Sonehalf}$ and $\ket{\Dfivehalf}$ of the three level system.
The expected contrast $C$ and phase $\Phi$ can then be written as
\begin{equation}
 C = \sqrt{\Tr^2(\sigma_x^\text{SD} \rho) + \Tr^2(\sigma_y^\text{SD} \rho)},
 \label{eq:contrast}
\end{equation}
\begin{equation}
 \Phi = \arg\left[\Tr(\sigma_x^\text{SD} \rho) + i \Tr(\sigma_y^\text{SD} \rho)\right].
 \label{eq:phase}
\end{equation}

Measuring $P_\text{D$_{5/2}$}$ as a function of the phase $\phi'$ of the second
$\frac{\pi}{2}$-pulse 
allows us to fit a sinusoidal curve to the data and extract contrast and phase
from the fit.
If the contrast is equal to one the ion was still in a fully coherent superposition of the
\Sonehalf{} and \Dfivehalf{} states and there was no spontaneous decay of the
\Pthreehalf{} state.
If the contrast is zero all the initial \Sonehalf{} state population had been
transferred to the \Pthreehalf{} state and the inevitable spontaneous decay had
destroyed all coherence.
The phase on the other hand is expected to undergo jumps of value $\pi$ each
time the ion's state population is transferred to the \Pthreehalf{} state and
back again to the \Sonehalf{} state by the pulse train.
This is due to the fact that a two-level system picks up a sign after a
$2 \pi$ rotation 
$\ket{0} \overset{2 \pi}{\longrightarrow} -\ket{0}$.

\begin{figure*}
\centering
 \includegraphics[width=0.8\textwidth]{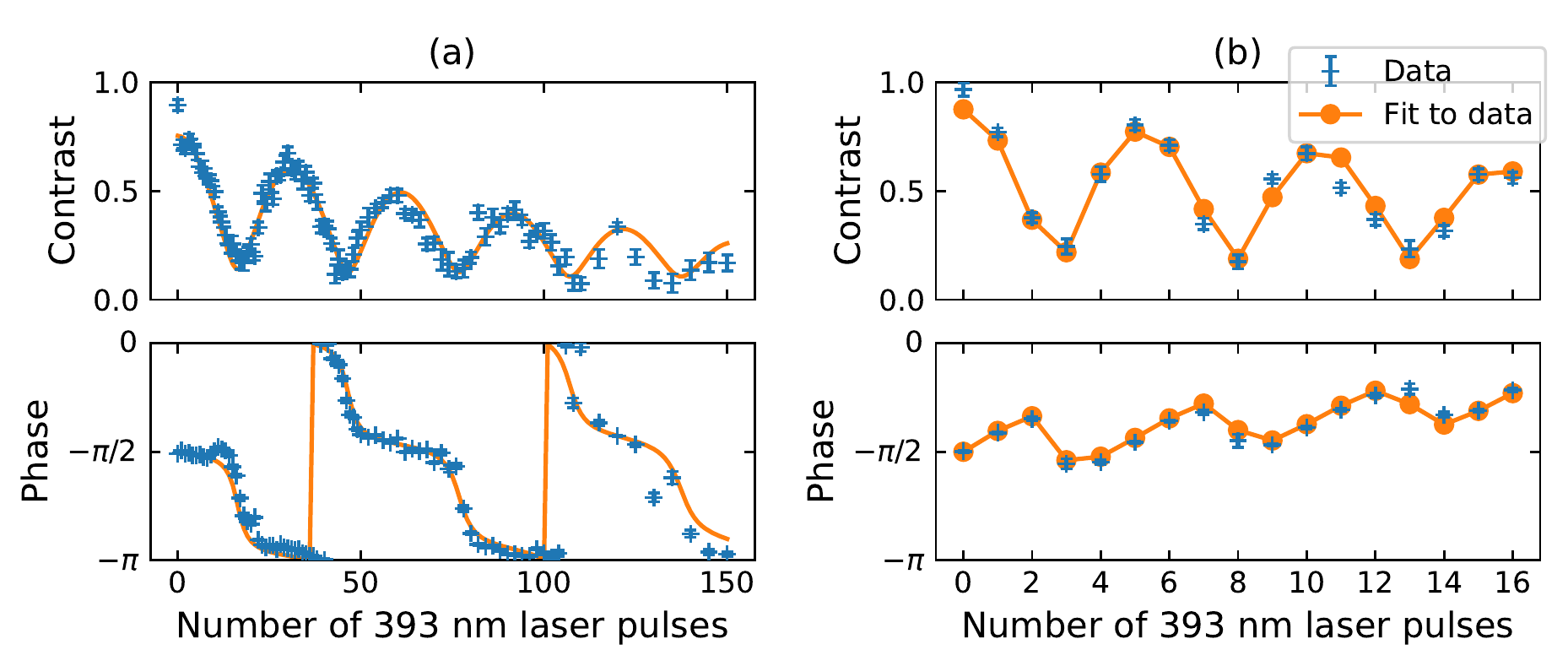}
 \caption{
  Contrast and phase of the experiment described in figure \ref{fig:ramsey_sequence}.
  Each pair of contrast/phase data points was obtained by a two-step process:
  For a given number of \SI{393}{\nm} laser pulses the experiment was repeated
  100 times for each of 41 different phases from \numrange{0}{4 \pi}. Then the
  contrast and phase were extracted from a fit to that data. This was repeated
  for each number of laser pulses.
  The solid circles and the connecting curve were computed using equations (\ref{eq:simulation}),
  (\ref{eq:contrast}) and (\ref{eq:phase}), and the parameters determined by a fit to the data.
  Additionally, the contrast is scaled by another fit parameter $C$ to 
  accommodate the reduced contrast, which we believe is caused by unwanted
  leaking light during the experiment.
  (a) Data set taken at a repetition rate of \SI{5}{\GHz} at low light power
  ($\theta = \SI{0.065}{\pi}$,
  $\Delta = \SI{0.03159}{\GHz}$,
  $\Delta^\prime = \SI{-0.003626}{\GHz}$,
  $C = 0.7541$).
  (b) Data set taken at \SI{1.25}{\GHz} repetition rate. We determine
  $\theta = \SI{0.377}{\pi}$,
  $\Delta = \SI{0.1310}{\GHz}$,
  $\Delta^\prime = \SI{-0.1866}{\GHz}$,
  $C = 0.8761$.
 }
 \label{fig:ramseys}
\end{figure*}

We repeat the measurement as a function
of the number $n$ of \SI{393}{\nm} laser pulses.
Figure \ref{fig:ramseys} (a) shows a data set where we plot the
remaining coherence and phase against the number of pulses.
For this set it took about 15 optical pulses at \SI{5}{\GHz} repetition rate
to complete a $\pi$-pulse,
which corresponds to $\theta \approx \SI{0.067}{\pi}$.
Nevertheless, one can already see that we can drive the
\trans{\Sonehalf}{\Pthreehalf} transition coherently and that each time we
return to the \Sonehalf{} state, the data is consistent with the observation of
phase jumps by $\pi$.

In figure \ref{fig:ramseys} (b) we plot the same quantities of another
data set taken at the same laser power as the
measurements presented in the other subsections.
We can observe the same structure as in the first data set but we need only about
2.5 optical pulses to complete one $\pi$-pulse on the transition.
A fit of the simulation of the ion-light interaction to the
experimental data allows us to extract $\theta = \SI{0.389+-0.005}{\pi}$.

\subsection{Discussion}
  \label{sec:discussion}

As shown above, we were able to deduce the rotation angle per pulse $\theta$ by letting an ion interact
with hundreds of consecutive pulses and afterwards measuring the probability
of finding the ion in the \Dfivehalf{} state.
Using only single pulses allowed us to measure not only $\theta$ again, but
also the phase offset of the first pulse after a dark time,
i.e. a pause in the pulse train.
By injecting single pulses into the waiting time between the two $\pi/2$-pulses
of a Ramsey experiment, we were able to gain information on how the pulses
influence the phase of the $\ket{\Sonehalf}$ state
as well as measure $\theta$.

\begin{table}
 \caption{Summary of fit parameters for the different experiments}
 \label{tab:fitsummary}
 \centering
 \begin{tabular}{clSS}
\firsthline \hline
{Rep. rate (\si{\GHz})} & Experiment &
{$\theta$ ($\pi$)} &
{$\delta \phi_{1,4}$ ($\pi$)}\\ 
\multirow{4}{*}{5} & interferometer &  & 0.74 { or } 1.26\\
& many pulses & 0.227+-0.012 & \\
& single pulses & 0.195+-0.018 & 1.282+-0.001\\ \hline 
\multirow{4}{*}{2.5} & interferometer &  & 0.37\\
& many pulses & 0.323+-0.005 & \\
& single pulses & 0.312+-0.010 & 0.361+-0.002\\ \hline 
\multirow{4}{*}{5/3} & interferometer &  & 0.10\\
& many pulses & 0.363+-0.005 & \\
& single pulses & 0.339+-0.007 & 0.088+-0.003\\ \hline 
\multirow{4}{*}{1.25} & interferometer &  & 0.04\\
& many pulses & 0.345+-0.005 & \\
& single pulses & 0.358+-0.003 & 0.051+-0.001\\
& Ramsey exp. & 0.377+-0.005 & \\
\hline \lasthline
 \end{tabular}
\end{table}

We now have three reliable ways to measure $\theta$
and the phase offset of the first pulses after a dark time.
The methods produce the same results within their respective error margins
but each have their advantages and disadvantages as described below.
Table \ref{tab:fitsummary} summarizes our results.


Using many pulses to pump into a dark state allows us to measure the laser-ion system's
rotation angle per pulse $\theta$ and requires neither the ability to pick single pulses nor
a laser resonant to the
\trans{\Sonehalf}{\Dfivehalf}, \SI{729}{\nm}, \SI{1}{\Hz} linewidth, quadrupole transition.
We therefore only need additional lasers that are readily available and
only need to be stabilized to a linewidth of $\lesssim\SI{1}{\MHz}$, which can be
achieved easily.
The only parameters that need to be known to fit the data are the number of
pulses $n$ and the well-known decay rates $\Gamma_\text{PS}$ and $\Gamma_\text{PD}$.
From the fit we can extract $\theta$ and also the
detuning between the transition frequency and a laser mode.
This allows us to tune the laser into resonance with the transition.

The method works well for any laser repetition rate $f_\text{rep}$ as long
as it is much larger than the \trans{\Sonehalf}{\Pthreehalf} transitions
linewidth.
Also, the rotation angle should satisfy $\theta \leq \pi$, since
rotation angles $\theta' = \pi + \delta$ can not be discerned from
$\theta = \pi - \delta$.


By using only single pulses to pump into a dark state (method 2) we can observe single
pulse dynamics and gain additional information about the pulse characteristics,
such as changing phase shifts and pulse powers.
With these measurements we are able to reproduce results obtained both with
long pulse trains and with the Michelson interferometer.


The Ramsey contrast technique (method 3) is experimentally more challenging than the
previous procedures.
We need to be able to create a coherent superposition of the \Sonehalf{} and
\Dfivehalf{} states which requires a few-\si{\kHz} linewidth laser.
It is also necessary that any AC Stark shift has been canceled, otherwise
it is not possible to transfer all \Sonehalf{} state population to the
\Pthreehalf{} state as required.
The advantage is that we can immediately see if we have reached our goal and if
one pulse suffices to do a $\pi$-pulse on the ion.
By fitting an empirical model to the data we can also learn how many pulses
are currently needed to do a $\pi$-pulse, if the power is not sufficient, yet.
Furthermore, the method allows one to track the phase of the population
in the \Sonehalf{} state because the \Dfivehalf{} state population serves as a
phase reference.

A fourth method \cite{Zanon-Willette2011} exists that we have not pursued
systematically.
It is based on measuring the fluorescence rate of the ion while it is
interacting with a continuous pulse train.
While it is experimentally easy to conduct, it requires precise knowledge of
the detector's total efficiency.

As stated before, almost all experiments
presented in this work were conducted at an average output power of the
high power EDFA of \SI{\approx 600}{\mW} which corresponds to
\SI{\approx 10}{\mW} \SI{393}{\nm} light power.
Nevertheless, we are able to increase the \SI{393}{\nm} light power to
$\gtrsim \SI{100}{\mW}$ as seen in figure \ref{fig:conversion_efficiency},
which suffices to create pulses that act as $\pi$-pulses with
\SI{96.4+-1.9}{\%} probability.

\section{Conclusion}
  \label{sec:conclusion}

In summary, we have designed, set up and characterized a high repetition rate laser,
derived from a stabilized optical frequency comb, which is suitable for
the implementation of ultrafast quantum gate operations with trapped \Ca{} ions.
We amplify the light at
\SI{1572}{\nm} and shift the wavelength via cascaded SHG to \SI{393}{\nm},
resonant to the \trans{\Sonehalf}{\Pthreehalf} transition in \Ca{}.
We have demonstrated that we can pick arbitrary pulse sequences out of the
\SI{5}{\GHz} pulse train and that our laser can coherently drive the
\trans{\Sonehalf}{\Pthreehalf} transition in \Ca{}.

We have developed and applied three different techniques to measure
the rotation angle per pulse $\theta = \Omega \, \delta t$
of a pulsed laser-ion system and
shown that we can create approximate $\pi$-pulses on the
\trans{\Sonehalf}{\Pthreehalf} transition with only a single optical pulse:
a key-requisite to implementing a resonant, ultrafast, two-qubit phase gate operation.


\section*{References} 
\providecommand{\newblock}{}

\end{document}